\documentclass[twocolumn]{aastex62}

\usepackage{graphicx}	
\usepackage{amsmath}	

\newcommand{\solar}{$_\odot$}
\newcommand{\zml}{$Z = 10^{-4}$}

\newcommand{\solarperyr}{$_\odot \, \textrm{yr}^{-1}$}

\newcommand{\binaryc}{\texttt{binary\_c}}

\newcommand{\Mwd}{$M_{\rm WD}$}

\newcommand{\Mig}{$M_{\rm ig}$}
\newcommand{\Mdot}{$\dot{M}$}

\newcommand{\Z}{$Z$}
\newcommand{\tento}[1]{$10^{#1}$}
\newcommand{\timestento}[2]{$#1 \times 10^{#2}$}
\newcommand{\iso}[2]{\hbox{${}^{#1}{\rm #2}$}}

\received{January 1, 2018}
\revised{January 7, 2018}
\accepted{\today}
\submitjournal{ApJ}

\shorttitle{Novae and Galactic Li}
\shortauthors{Kemp et al.}

\begin{document}

\title{Viability of novae as sources of Galactic lithium}

\correspondingauthor{Alex Kemp}
\email{alexander.kemp@monash.edu,}

\author[0000-0003-2059-5841]{Alex J. Kemp}
\affil{School of Physics \& Astronomy, Monash University, Clayton 3800, Victoria, Australia}
\affil{Centre of Excellence for Astrophysics in Three Dimensions (ASTRO-3D), Melbourne, Victoria, Australia}

\author{Amanda I. Karakas}
\affil{School of Physics \& Astronomy, Monash University, Clayton 3800, Victoria, Australia}
\affil{Centre of Excellence for Astrophysics in Three Dimensions (ASTRO-3D), Melbourne, Victoria, Australia}

\author{Andrew R. Casey}
\affil{School of Physics \& Astronomy, Monash University, Clayton 3800, Victoria, Australia}
\affil{Centre of Excellence for Astrophysics in Three Dimensions (ASTRO-3D), Melbourne, Victoria, Australia}

\author{Benoit C\^{o}t\'{e}}
\affil{Department of Physics and Astronomy, University of Victoria, Victoria, BC V8P 5C2, Canada}
\affil{Konkoly Observatory, Research Centre for Astronomy and Earth Sciences, E\"otv\"os Lor\'and Research Network (ELKH), Konkoly Thege Mikl\'{o}s \'{u}t 15-17, H-1121 Budapest, Hungary}

\author{Robert G. Izzard}
\affil{Astrophysics Research Group, University of Surrey, Guildford, Surrey GU2 7XH, UK}

\author{Zara Osborn}
\affil{School of Physics \& Astronomy, Monash University, Clayton 3800, Victoria, Australia}
\affil{Centre of Excellence for Astrophysics in Three Dimensions (ASTRO-3D), Melbourne, Victoria, Australia}

\begin{abstract}

Of all the light elements, the evolution of lithium (Li) in the Milky Way is perhaps the most difficult to explain. Li is difficult to synthesize and is easily destroyed, making most stellar sites unsuitable for producing Li in sufficient quantities to account for the proto-solar abundance. For decades, novae have been proposed as a potential explanation to this `Galactic Li problem', and the recent detection of \iso{7}Be in the ejecta of multiple nova eruptions has breathed new life into this theory. In this work, we assess the viability of novae as dominant producers of Li in the Milky Way.
We present the most comprehensive treatment of novae in a galactic chemical evolution code to date, testing theoretical- and observationally-derived nova Li yields by integrating metallicity-dependent nova ejecta profiles computed using the binary population synthesis code \binaryc\ with the galactic chemical evolution code \texttt{OMEGA+}.
We find that our galactic chemical evolution models which use observationally-derived Li yields account for the proto-solar Li abundance very well, while models relying on theoretical nova yields cannot reproduce the proto-solar observation. A brief exploration of physical uncertainties including single-stellar yields, the metallicity resolution of our nova treatment, common-envelope physics, and nova accretion efficiencies indicates that this result is robust to physical assumptions. Scatter within the observationally-derived Li yields in novae is identified as the primary source of uncertainty, motivating further observations of \iso{7}Be in nova ejecta. 
\end{abstract}


\section{Introduction} \label{sec:intro}




Lithium (Li) is a notoriously troublesome element. It is extremely fragile, destroyed in H-capture reactions at temperatures as low as \timestento{2}{6} K. This is cool enough for stars to deplete their surface Li on, or before, the main sequence by convecting surface material deeper into the star where the Li is destroyed. This makes it almost impossible to accurately calculate the Li abundance at birth for most stars. The exception to this is the Sun, for which we have access to meteorites which preserve the proto-solar Li abundance  \citep[\textit{A}(Li)=$3.26\pm0.05$, see][]{lodders2009,asplund2009}. Galactic chemical evolution (GCE) models typically under-produce this Li abundance by roughly an order of magnitude, a discrepancy known as the `Galactic Li problem' \citep{dantona1991,matteucci1995}.

The fragility of Li is detrimental to its production. Its dominant isotope \iso{7}Li is formed as the sole decay product of \iso{7}Be, which is unstable with a half-life of 53.3 days. \iso{7}Be forms via the proton-proton (pp) chains during H-burning, but quickly decays to \iso{7}Li via electron capture (pp II). The \iso{7}Li itself is then also destroyed by proton capture, with the net result that very little survives H-burning. The Cameron-Fowler mechanism \citep{cameron1971} proposes that some of the \iso{7}Be is transported to a cooler region, where electron captures and the natural decay of \iso{7}Be can form Li in an environment where it can survive. 

Although the yield is model dependent, relatively little Li can be produced in this way by AGB stars \citep[e.g.,][]{karakas2010,pignatari2016}, and core-collapse supernova models produce orders of magnitude less \citep[e.g.,][]{kobayashi2006}. Li can also be produced through spallation reactions caused by cosmic rays at levels comparable to AGB stars \citep{prantzos2012}. 

The list of viable Li sources at this point is short but not exhausted. It has been proposed for many years now that classical novae could be viable sites for Li production \citep{arnould1975}. Novae are transients caused by explosive H-burning episodes on the surface of accreting white dwarfs (WDs). Their high burning temperatures, rapid evolution, and high mass loss rates during outburst make them promising Li producers due to their ability to synthesize \iso{7}Be in significant quantities and then transport that material into cooler mass-losing regions during outburst, where it can decay to \iso{7}Li without that Li being destroyed. Early theoretical work on explosive H burning and nova yields found significant overproduction of Li relative to solar composition material \citep{arnould1975,starrfield1978}. More sophisticated modelling with comprehensive reaction networks qualitatively confirmed these early results \citep{jose1998,starrfield2000}, solidifying the idea of novae as Li factories.

Since the first detection of \iso{7}Be in 2015 \citep{tajitsu2015}, \iso{7}Be has been been detected in the spectra of a handful of nova outbursts \citep{tajitsu2015,molaro2016,tajitsu2016,selvelli2018,molaro2020,arai2021,izzo2022}. These observations not only provided direct evidence of novae producing Li, but also implied a far greater over-production factor than that found in theoretical models, reigniting interest in novae as a solution to the Galactic Li problem. Where older chemical evolution models that relied on theoretical nova yields identified novae as promising candidates \citep{romano2001}, newer works making use of the observationally-derived yields demonstrated that novae could account for the high proto-solar Li abundance \citep{cescutti2019,grisoni2019}.

However, existing calculations on the viability of novae as dominant sources of Li have relied upon simplified models for how novae behave. For example, the recent work of \cite{grisoni2019} assume each nova system to undergo \tento{4} eruptions \citep{bath1978}, assigning each a constant ejecta mass of Li bounded by observations of V1369 Cen \citep{izzo2015}. The birth rate of nova systems is set as a fraction of the WD formation rate, and all nova eruptions generated from a period of star formation are assumed to occur instantaneously, offset by a fixed amount time. \cite{cescutti2019} features a more sophisticated treatment which approximates the continuous release of nova ejecta, and instead treats the average Li ejecta-mass per event as a free parameter, ultimately determined to be consistent with observations of \iso{7}Be in nova ejecta. These assumptions were necessary as more sophisticated models for nova populations did not exist.

In this work, we instead rely on the results of previously computed theoretical nova populations \citep{kemp2022}, where each nova eruption is treated individually based the white dwarf mass \Mwd\ and accretion rate \Mdot. Our chemical evolution model and treatment of novae is described in Section \ref{sec:methodology}, our findings on novae as sources of Li in the Milky Way presented Section \ref{sec:results}, and commentary on the significance and uncertainty in our findings presented in Section \ref{sec:discussion}. We conclude in Section \ref{sec:conclusion}. Supplementary material can be accessed here: \href{https://doi.org/10.5281/zenodo.6644271}{DOI:10.5281/zenodo.6644271}

\section{Methodology}
\label{sec:methodology}
\subsection{Nova treatment}

\begin{figure}
\epsscale{0.8}
\plotone{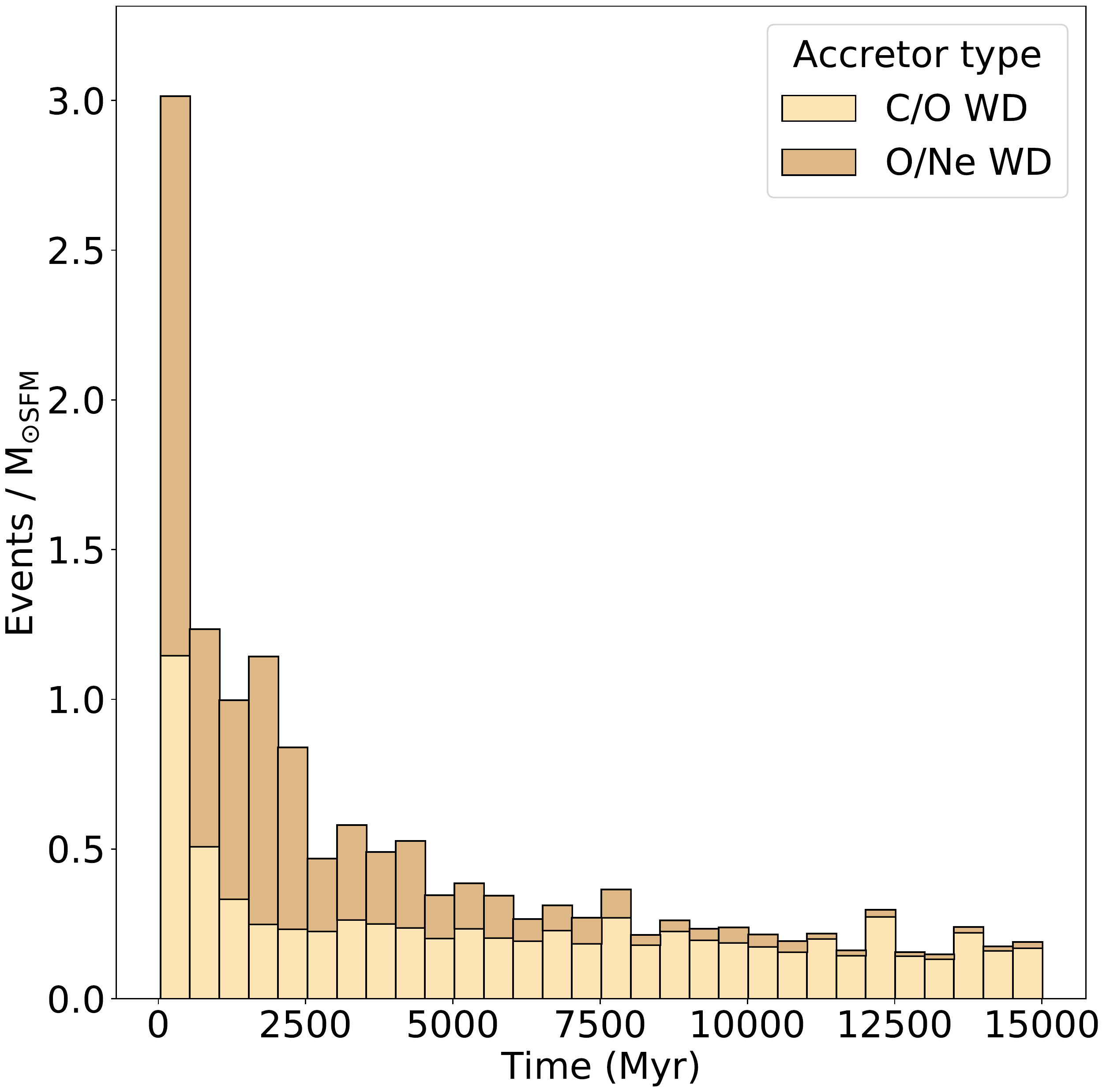}
\plotone{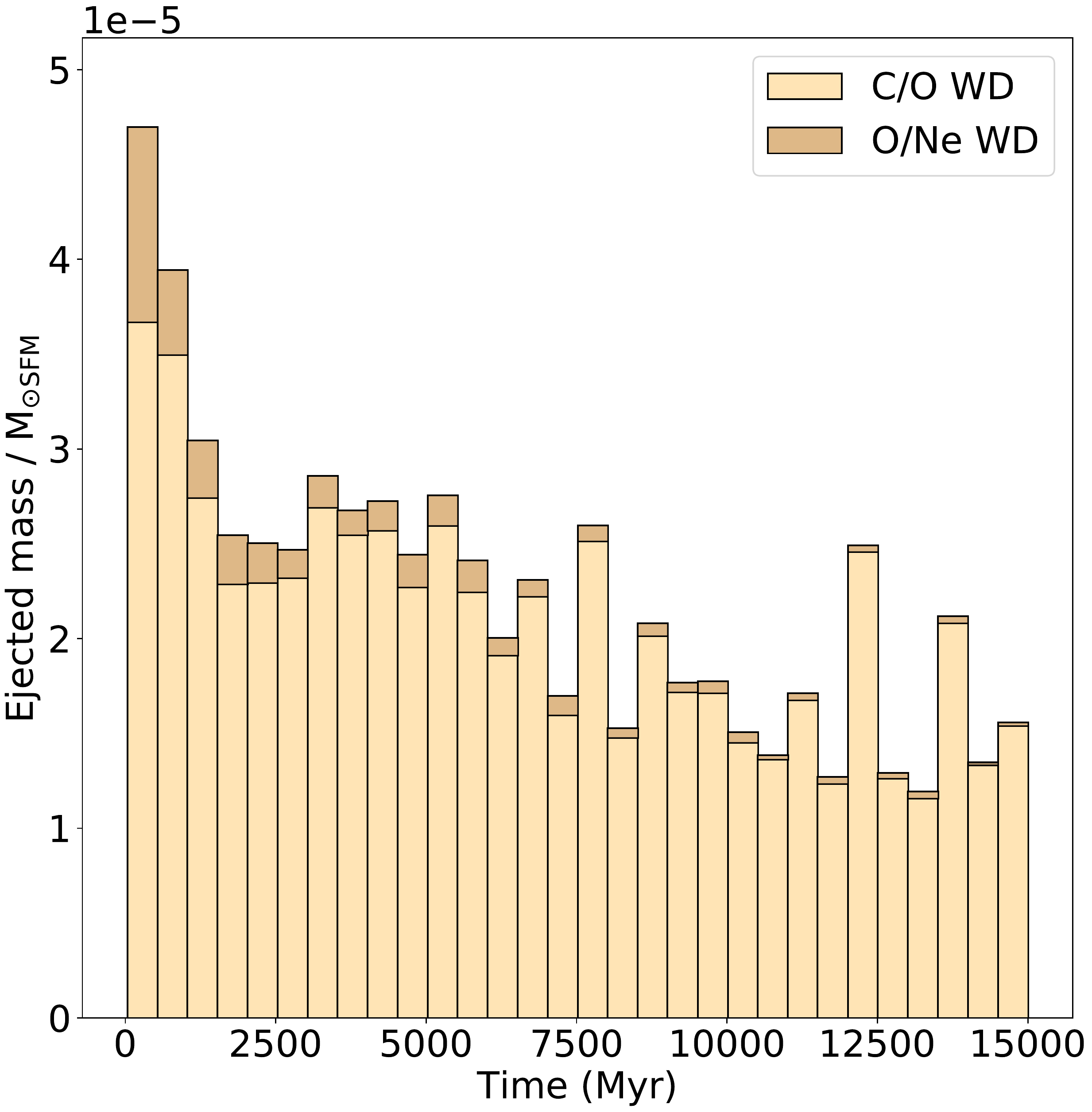}
\caption{Delay time (upper panel, tracking when the nova eruptions occur relative to star formation) and ejecta-delay time (lower panel, tracking when mass is ejected relative to star formation) distributions for \Z=\tento{-3}. Both distributions are normalised per solar mass of star forming material $\rm M_{\rm \odot SFM}$, \citep[for details, see][]{kemp2021,kemp2022}.}
\label{fig:dtds}
\end{figure}

Figure \ref{fig:dtds} shows a nova delay time (upper panel) and ejecta-delay time (lower panel) distribution coloured by the WD composition. The delay time distribution tracks the time from star formation to each nova eruption, where each nova eruption is weighted equally. The ejecta-delay time distribution instead weights each nova eruption by the total mass ejected into the interstellar medium, tracking the mass ejection profile of novae for each burst of star formation.

A metallicity-dependent grid of these pre-computed ejecta-delay time distributions is taken as a direct input into our galactic chemical evolution models. The underlying binary population synthesis models are a subset (\zml, \tento{-3}, \timestento{5}{-3}, 0.01, 0.02) of those presented in \cite{kemp2022}, and further details about the process of simulating each nova eruption in \binaryc\ \citep{izzard2004,izzard2006,izzard2009,izzard2018binary} are available in \cite{kemp2021}.

To make predictions about specific elements or isotopes, these ejecta-delay time distributions must be combined with nova yield tables. We can divide our nova population according to the white dwarf mass \Mwd\ and accretion rate \Mdot\ at the time of eruption by using an array of ejecta-delay time distributions that correspond to different regions of \Mwd-\Mdot parameter space. This allows us to map physics-dependent yields from theoretical models \citep{jose1998,starrfield2009,starrfield2020,rukeya2017,jose2020} to their relevant ejecta-delay time distributions. This treatment innately accounts for ejecta mass and delay time variation in different nova systems, the metallicity dependence of nova-system formation and evolution, and physics-dependant aspects of nova nucleosynthesis \citep[][]{iliadis2015}.

\subsection{Nova yield sets}

We have compiled five theoretical nova nucleosynthesis yield profiles. The implementation of each of these yield profiles is described below:

\begin{enumerate}
    \item J1998: Makes use of 50\% pre-enriched C/O WD nucleosynthesis yields for \Mwd\ = 0.8 and 1.0 M\solar, and 50\% pre-enriched O/Ne WD yields for \Mwd\ = 1.15, 1.25, and 1.35 M\solar\ \citep{jose1998}.
    
    \item S2009/2020: Makes use of 50\% pre-enriched C/O WD nucleosynthesis yields for \Mwd\ = 0.6, 0.8, 1.0 and 1.15 M\solar\ \citep{starrfield2020}, and 50\% pre-enriched O/Ne WD yields for \Mwd\ = 1.25, 1.35 M\solar\ \citep{starrfield2009}.
    
    \item J2020: Makes use of C/O WD nucleosynthesis yields for \Mwd\ = 1.0 M\solar\ and O/Ne WD yields for \Mwd\ = 1.15 M\solar\ \citep{jose2020}. Rather than assuming a pre-enrichment fraction to account for mixing during the eruption, these yields come from models which instead simulate the eruption by combining 3D \citep[\texttt{FLASH};][]{fryxell2000} and 1D \citep[\texttt{SHIVA};][]{jose1998} methods to model the conditions at outburst.
    
    \item R2017: Makes use of 50 percent pre-enriched Li yields for C/O WDs at \Mwd\ = 0.51, 0.6, 0.7, 0.8, 0.9, 1.0, and 50\% pre-enriched O/Ne WD yields for \Mwd\ = 1.1, 1.2, 1.3, and 1.34 M\solar\ \citep{rukeya2017}. This yield profile is notable for also providing yields for a range of accretion rates for each \Mwd, rather than sampling only at a fixed accretion rate of \timestento{2}{-10} M\solarperyr\ as in the previously described J1998, S2009/2020, and J2020 sets. A figure demonstrating the breakdown of this grid is provided in the supplementary material.
    
    \item R2017simple: Makes use of 50\% pre-enriched Li yields for C/O WDs at \Mwd =  0.8 and 1.0 M\solar, and O/Ne WD yields for \Mwd = 1.15, 1.25, and 1.35 M\solar\ \citep{rukeya2017}. This yield set is, by design, identical to the parameter space set out in the J1998 set. It is useful as it can be directly compared with the R2017 set to assess the impact of only sampling the accretion rate space at \timestento{2}{-10} M\solarperyr.

\end{enumerate}

Note that all of the underlying models informing the above theoretical yield profiles assume solar composition material.

\begin{figure*}
\epsscale{1}
\plotone{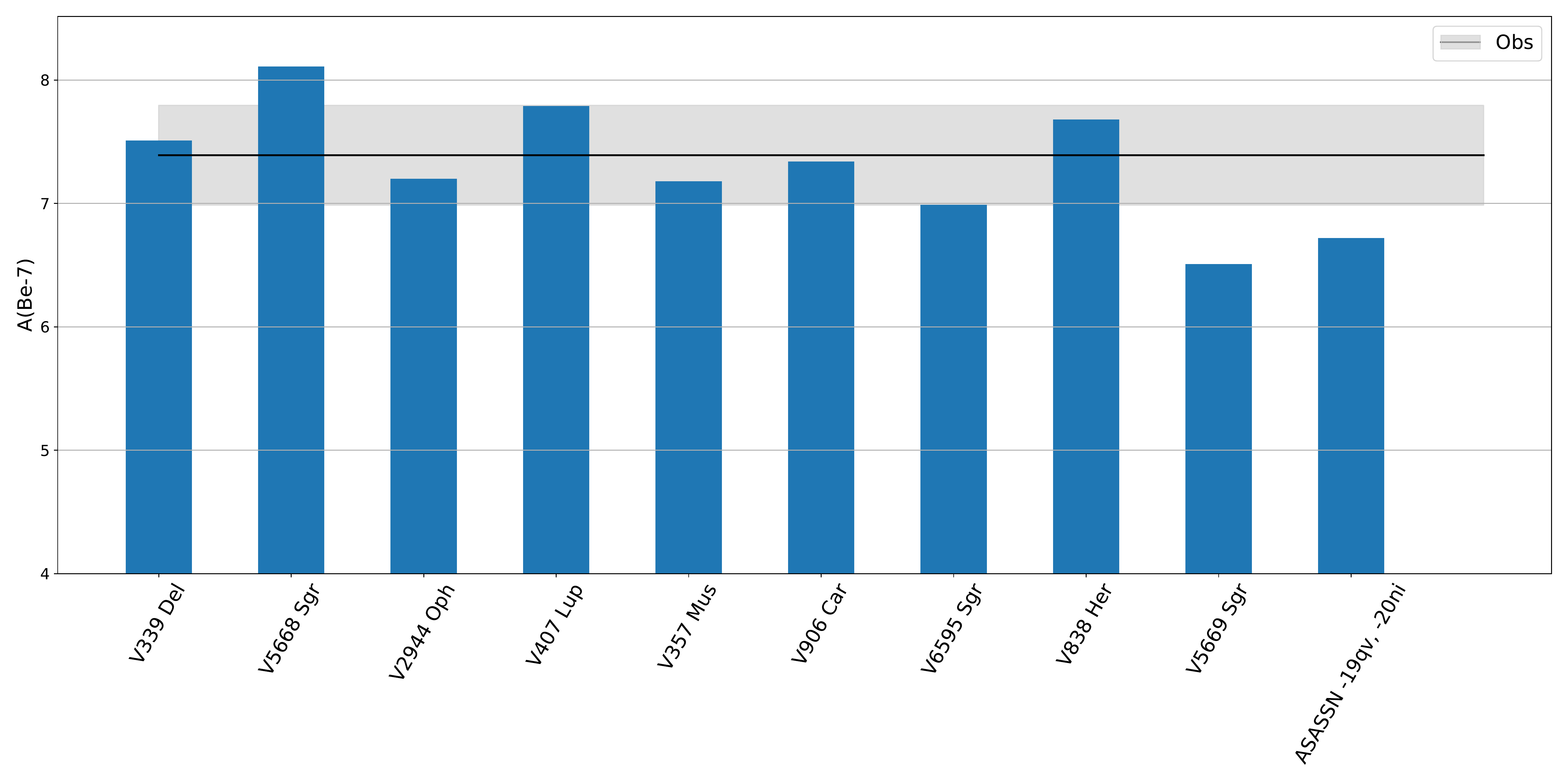}
\caption{Observationally-derived \textit{A}(\iso{7}Be), corrected for \iso{7}Be decay since eruption. Decay-corrected \textit{A}(\iso{7}Be) values are taken from: \cite{molaro2020} table 1 (V339 Del, V5668 Sgr, V2944 Oph, V407 Lup,
V357 Mus, V906 Car),
\cite{molaro2022} table 4 (V6595 Sgr, V838 Her, V5669 Sgr), and
\cite{izzo2022} (ASASSN-19qv, ASASSN-20ni).
}
\label{fig:obs}
\end{figure*}

Figure \ref{fig:obs} presents all available observationally-derived \iso{7}Be abundances \textit{A}(\iso{7}Be), corrected for \iso{7}Be decay \citep{molaro2020,molaro2022}. Over-plotted are the mean and 1-$\sigma$ error bars for this data computed using all available observations (\textit{A}(\iso{7}Be) = $7.12 \pm 0.71$, grey). The H mass fraction $X$ of the ejecta is required to convert this information into a useful mass fraction of \iso{7}Be, a quantity dependent on the fraction of core material mixed into the burning zone, which we must assume. We assess the impact of this assumption by considering two cases for the mixing fraction, 50\% ($X=0.35$) and 25\% ($X=0.5$), with the associated $X$ values for these two cases based on the simulations of \cite{jose1998} and \cite{starrfield2009,starrfield2020}.

Significant scatter exists in these measurements. The novae V407 Lup, V6595 Sgr, and V838 Her are classified as O/Ne novae due to the detection of bright Ne lines in their late nebula phases \citep[table 4,][]{molaro2022}. Of these novae, only V407 Lup and V838 Her have measurements notably above average. The dusty nova V5668 Sgr has the highest value (\textit{A}(\iso{7}Be$) = 8.1$), but V357 Mus and V906 Car appear typical despite also being dusty \citep[][and references therein]{gordon2021}.



\subsection{Milky Way chemical evolution model}

Our chemical evolution model is computed using the two-zone GCE code \texttt{OMEGA+} \citep{cote2018omegaplus}. The model structure consists of a star-forming galactic component coupled with a circum-galactic halo which functions as a hot gas reservoir and facilitates galactic recycling. The central galactic component is simulated using \texttt{OMEGA} \citep{cote2017omega}, a one-zone GCE code accounting for the chemical yields of different stellar populations.

Our baseline Milky Way model, to which we add nova contributions, uses an exponential Galactic inflow rate and calculates outflows as a function of star formation. Stellar yields for type Ia supernovae, asymptotic giant branch (AGB), and massive stars are taken from \cite{thielemann1986}, \cite{karakas2010}, and \cite{kobayashi2006} respectively. Contributions to Li production from cosmic rays are approximated as a function of [Fe/H] using data from Figure 16 of \cite{prantzos2012}.

Further details and figures used to validate our baseline Milky Way model can be found in the supplementary material.



\section{Results}
\label{sec:results}

\subsection{Lithium source comparison}

\begin{figure}
\centering
\includegraphics[width=1.0\columnwidth]{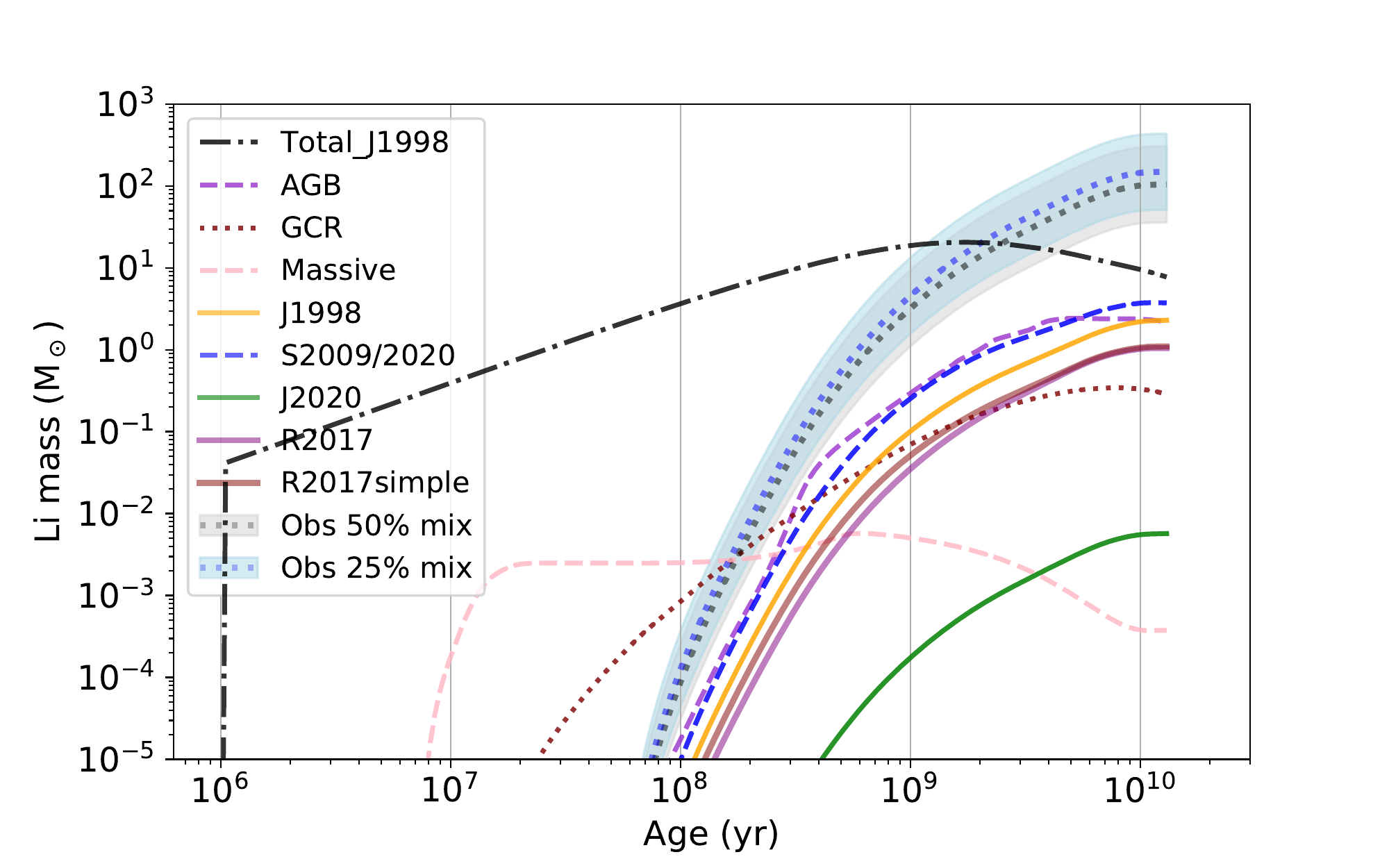}
\caption{Li mass contributions from different sources as a function of Galactic age. See main text for discussion.}
\label{fig:Limass_difsources}
\end{figure}

Figure \ref{fig:Limass_difsources} presents the mass of Li produced by different sources as a function of Galactic age. The total mass of Li for the full J1998 GCE model is shown for comparison, which includes primordial material being introduced through galactic inflows as well as AGB, massive stellar, Galactic cosmic ray (GCR), and nova contributions using the previously described J1998 yield profile.

Two different models relying on different observationally-derived yield sets are shown (dotted lines + shaded 1-$\sigma$ error bar): assuming 50\% mixing of underlying core material (grey); and assuming 25\% mixing (blue). Reducing the mixing fraction from 50\% to 25\% can be seen to increase the average noticeably, although the change is small relative to the observational uncertainty. Despite the large error bars resulting from the scatter in observational Li determinations, it is clear that even in the most pessimistic case novae are expected to produce far more Li than any other stellar source over the Galaxy's lifetime.

The S2009/2020 yield set results in the most Li produced by novae when theoretical nova yields are used, but even this yield set produces almost an order of magnitude less Li by $t=10$ Gyr than the lowest error bound of the observational yield sets. The disagreement between theoretical models of Li production in novae and observations of Li in the ejecta is at least a factor of 5.

Comparing the different theoretical nova yield profiles, the profiles relying on pre-mixing -- J1998, S2009/2020, R2017, R2017simple -- have broadly comparable productivity, overtaking GCR contributions after roughly 2 Gyr of evolution but never overproducing relative to AGB yields by more than a factor of 2 (S2009/2020). Note that the R2017 and R2017simple models are almost identical in Figure \ref{fig:Limass_difsources}, but there is discernible variation between the Li masses produced by yield profiles from different research groups. This strongly implies that neglecting to resolve the accretion-rate dependency of nova nucleosynthesis is of secondary importance. A figure presenting the distribution of Li mass production in \Mwd-\Mdot\ space according to the R2017 model is included in the supplementary material.

The only theoretical yield set that does not rely on pre-mixing -- J2020 -- produces three orders of magnitude less Li than the theoretical pre-enriched yield sets. 
\cite{jose2020} attribute this to the longer timescale over which the thermonuclear runaway develops in these models, which allows far more \iso{7}Be to be destroyed before it can be advected to cooler regions. A more detailed discussion is presented in \cite{jose2020}.


\subsection{Lithium in the Milky Way}

\begin{figure}
\centering
\includegraphics[width=1.0\columnwidth]{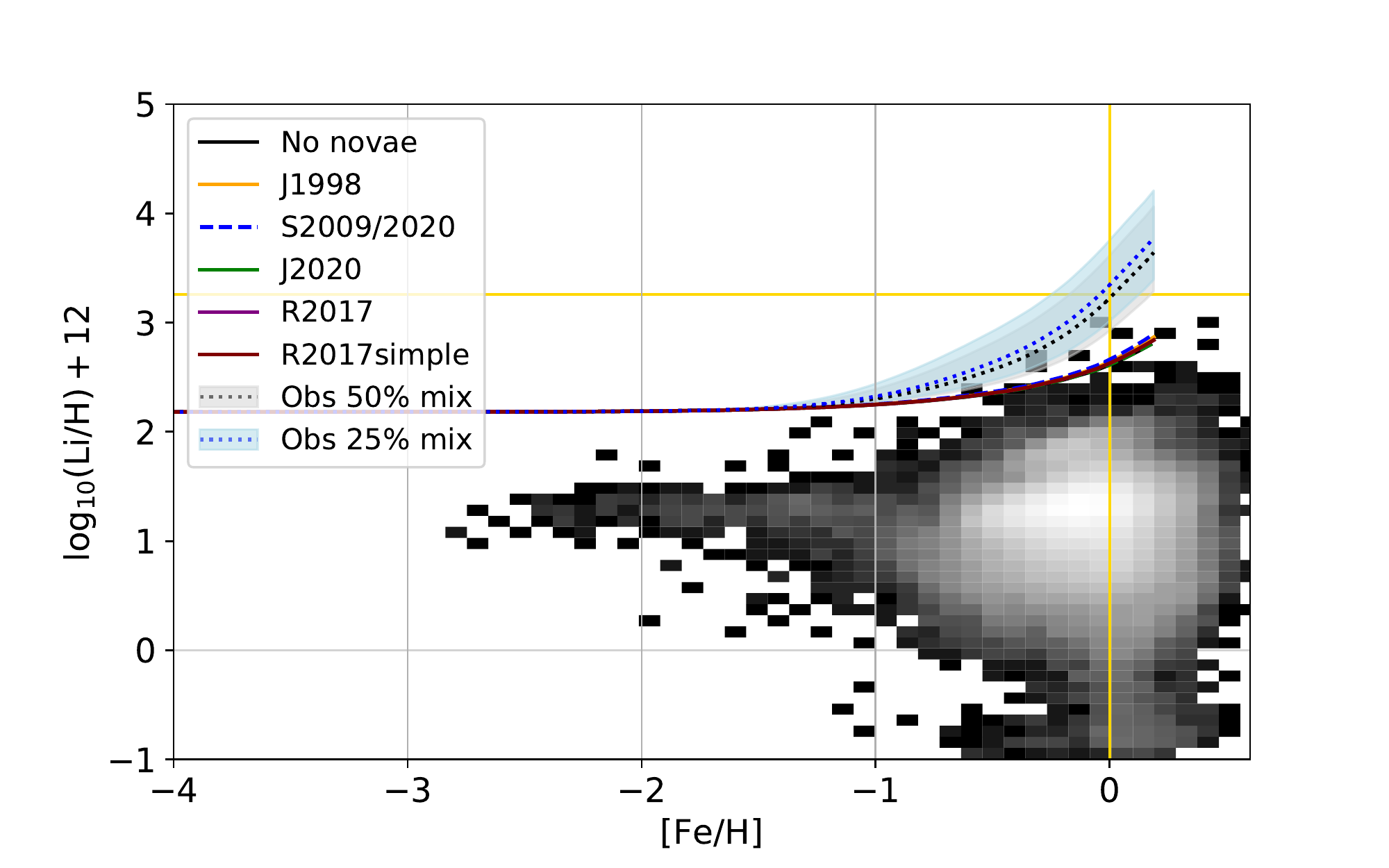}
\caption{Li abundance versus [Fe/H] in our Galactic models, overlaid with the proto-solar Li abundance \citep{lodders2009} (yellow cross-hair) and Li abundance data from GALAH DR3 \citep{buder2021}. All models include asymptotic giant branch, massive stellar, and galactic cosmic ray sources in addition to the designated nova yield set.}
\label{fig:LiAbu_GCR}
\end{figure}

Figure \ref{fig:LiAbu_GCR} presents the Li abundance versus [Fe/H] in each of our Galactic models, overlaid with the proto-solar Li abundance (yellow) and Li abundances of stars in GALAH DR3 \citep[][grey-scale]{buder2021}. The GALAH dataset used is comprised of the 85490 main sequence stars (\texttt{logg} $ > 3.7$) that had acceptable signal-to-noise  ratios (\texttt{snr\_c3\_iraf} $ > 30$), stellar parameters (\texttt{flag\_sp} $ = 0$), and Fe/H (\texttt{flag\_fe\_h} $ = 0$) and Li/Fe (\texttt{flag\_Li\_fe} $ = 0$) abundances. Unlike Figure \ref{fig:Limass_difsources}, which only plots the contributions specifically from novae in the curves labelled with nova yield profiles, each of the models in Figure \ref{fig:LiAbu_GCR} includes the specified nova yield profile in addition to all non-nova Li sources.

Figure \ref{fig:LiAbu_GCR} demonstrates that models making use of observationally-derived nova Li yields can account for the proto-solar Li abundance. Our model relying upon observational Li yield observational yield with the 50\% mixing assumption passes cleanly through the proto-solar Li observation. The Li mass fraction for novae used in this GCE model is approximately \timestento{4.8}{-5}, which we present as our empirical solution for the average Li mass fraction in nova ejecta. However, note that the model assuming 25\%  mixing produces a Li abundance at [Fe/H] = 0 that lies within 0.1 dex of the proto-solar observation, well within the 1-$\sigma$ uncertainties accounting for scatter in the observationally-derived \textit{A}(\iso{7}Be) data.

Conversely, all models relying on theoretical yield profiles are indistinguishable from each other or the baseline model without novae. These predict Li abundances an order of magnitude below the proto-solar Li abundance. Despite being non-negligible Li producers (Figure \ref{fig:Limass_difsources}), according to these yield-profiles novae simply do not produce enough Li to account for the solar Li abundance or, to a lesser extent, the upper envelope of the GALAH data.

It has been proposed \citep{grisoni2019} that the metallicity-dependence of novae could explain observations of declining Li abundances at high metallicities, a feature detected in the Milky Way by multiple groups \citep{mena2015,guiglion2016,bensby2018, fu2018,buder2018,buder2021}. We report that none of our models show signs of a Li decline at high metallicity, despite accounting for metallicity-dependent variation in our nova ejecta-delay time distributions.


\section{Discussion}
\label{sec:discussion}

GCE models incorporate, albeit often indirectly, a vast amount of physics. The adopted yield sets and productivities of different stellar phenomena (eg. AGB stars, massive stars, type Ia supernovae, neutron star mergers, etc.) all affect how a simulated galaxy evolves, and each of these prescriptions has its own physical assumptions. Additionally, modelling choices for galactic processes such as the star formation efficiency and inflow and outflow rates are vital. This modelling complexity renders assessment of uncertainty difficult.

Our baseline model of the Milky Way satisfactorily reproduces a range of observables (see supplementary material for supporting figures), including key elemental abundance trends. This does not necessarily make the model correct; rather, it is intended to provide a representative baseline model with which we can assess the importance of novae. In this work, we are concerned with Li, an element notoriously sensitive to stellar modelling choices. The two classes of assumptions most likely to affect our results are the adopted non-nova stellar yields and the assumed binary physics behind our ejecta-delay time distributions. We summarise here the results of a brief exploration of both these uncertainties.

We find that changing the adopted AGB yields from \cite{karakas2010} yields to NuGrid yields  \citep{pignatari2016} has minimal effect on the ability of our models using observational nova yield profiles to reproduce the solar Li abundance, despite the NuGrid yields producing roughly twice as much Li from AGB stars by [Fe/H]=0. Assessing the extreme lower bound, removing all non-nova sources of Li results in a 0.25 dex negative shift relative to the proto-solar abundance. This is well within the 1-$\sigma$ bounds associated with the observational data. Preliminary results of an investigation into the statistical effects of binary interactions on solar metallicity AGB yields -- an effect neglected in contemporary GCE codes -- indicate at most a 15\% reduction in Li (Osborn et al., in prep). We therefore conclude that our results are likely robust against uncertainties in non-nova Li yields.

Removing the metallicity dependence of the nova ejecta-delay time distributions and instead only using ejecta-delay time distributions computed at $Z=0.02$ results in a 35\% reduction in Li production from novae. This reduces the Li abundance at [Fe/H]=0 predicted by our GCE models using observationally-derived yield sets by 0.3-0.4 dex. The proto-solar Li abundance remains comfortably within the 1-$\sigma$ error bars associated with the observationally-derived yield profiles. We therefore conclude that our results are robust to the metallicity-resolution of our array of nova ejecta-delay time distributions.

\begin{figure}
\epsscale{1.2}
\plotone{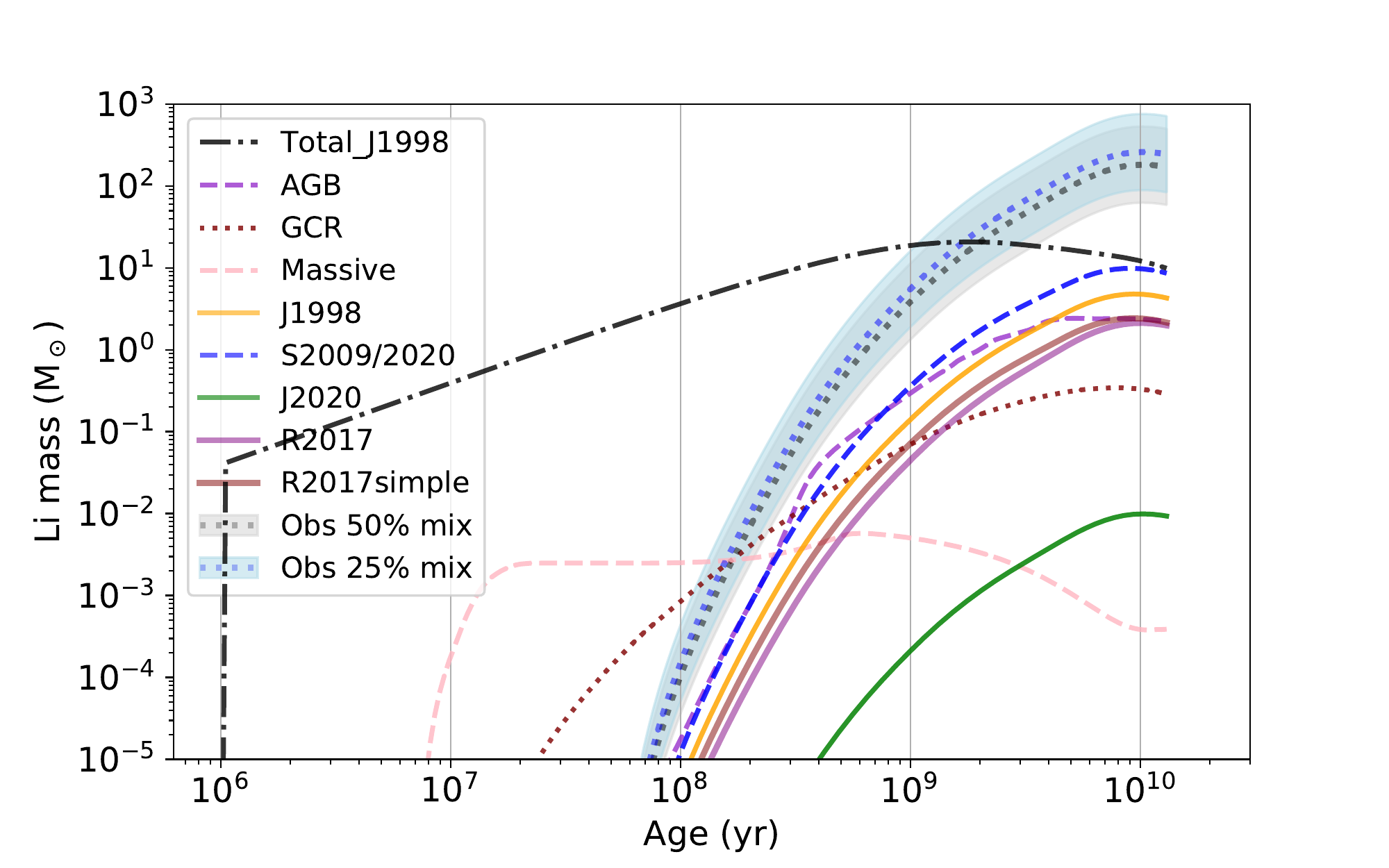}
\plotone{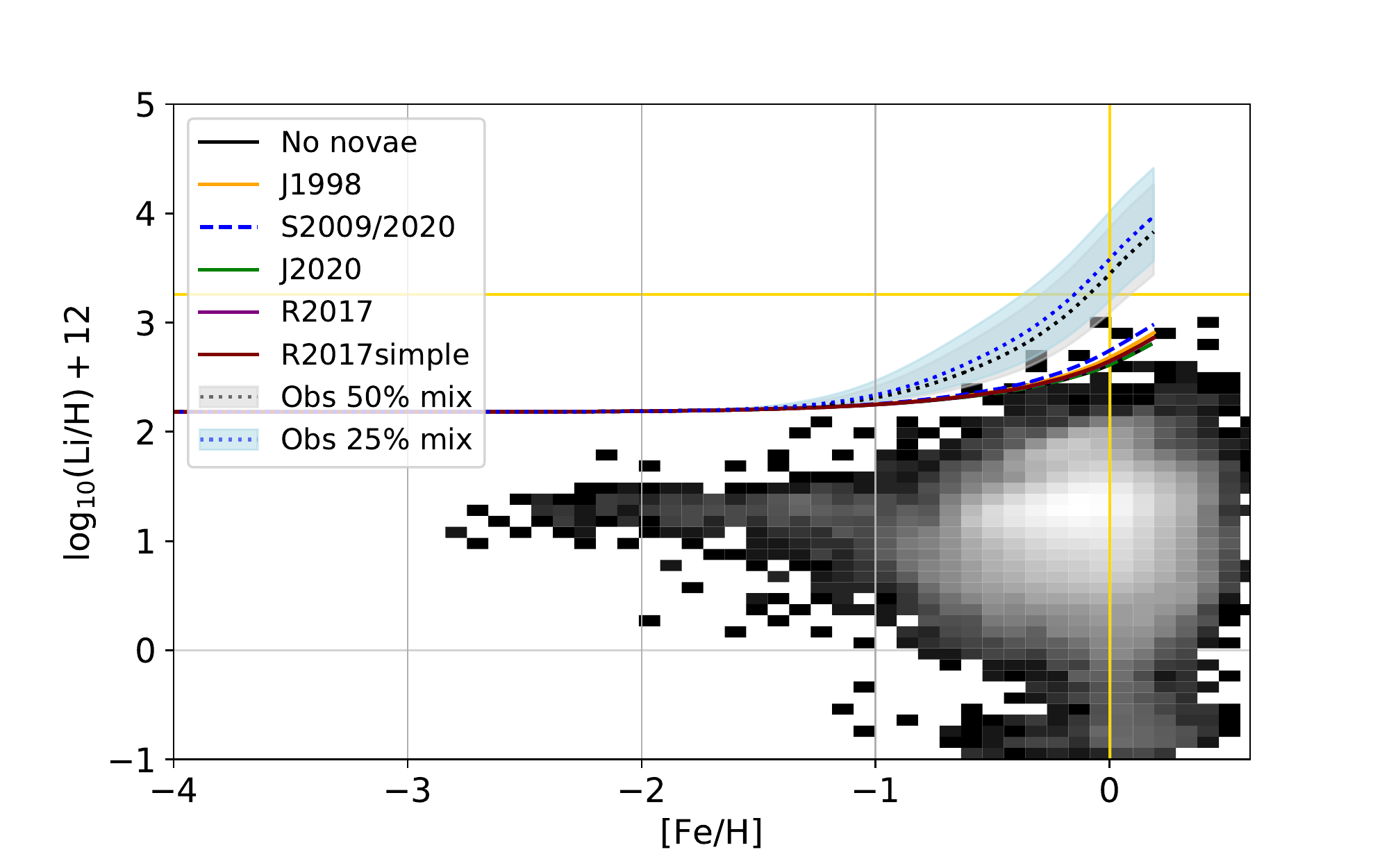}
\caption{As Figures \ref{fig:Limass_difsources} and \ref{fig:LiAbu_GCR} when common-envelope parameter $\lambda_{\rm CE}$ is set to 0.5, rather than the \cite{wang2016binding} prescription used in our baseline case. This physics case doubles the expected nova rate in M31  \cite[][]{kemp2021}, and we see this reflected in a doubling in Li production). The ability of observationally-derived nova yield profiles to reproduce the proto-solar Li abundance, and the inability of theoretical yield profiles, appears robust.}
\label{fig:lambda0p5}
\end{figure}

A full investigation of binary uncertainty is beyond the scope of this work. However, as an indicative example, Figure \ref{fig:lambda0p5} presents the results of using an alternative set of ejecta-delay time distributions computed using a different set of binary physics. In producing Figure \ref{fig:lambda0p5}, we have replaced the \cite{wang2016binding} prescription for the common-envelope parameter $\lambda_{\rm CE}$  \citep[][]{kemp2021} with a constant value of 0.5. This physics set was found in previous work \citep{kemp2021} to approximately double the predicted nova rate in M31, and represents an extreme case of variation due to binary physics \citep{kemp2021}. We find that the total mass of Li produced by novae approximately doubles under this assumption, implying roughly a factor of 2 uncertainty in our results from physical uncertainty in binary stellar physics.

We also investigated the effect of replacing the nova accretion efficiency prescription of \cite{wang2018} with a constant to assess the potential impact that nova-specific modelling choices could have on our results. Setting the accretion efficiency to 0.01, meaning that in each nova eruption 99\% of the accreted material is lost to the WD per eruption, has a negligible effect on our results.



\section{Conclusions}
\label{sec:conclusion}

In this work, we assess the importance of novae to the synthesis of Li in the Milky Way using the galactic chemical evolution code \texttt{OMEGA+}. Previous galactic chemical evolution models have either relied on simplified treatments of novae or neglected them altogether. In this work, we employ metallicity-dependent arrays of nova ejecta-delay time distributions computed using the binary population synthesis code \binaryc\ to model the Galactic nova population. 

We assess the viability of novae in the context of five different theoretical yield profiles \citep{jose1998,starrfield2009,rukeya2017,jose2020,starrfield2020} in addition to observationally-derived Li yields \citep{molaro2020,molaro2022,izzo2022}. We find that our Galactic chemical evolution models which make use of observationally-derived Li yields account for the proto-solar Li abundance very well, while the models which make use of theoretical yield profiles universally fail to reproduce this observation by an order of magnitude.

We find this result to be robust to all physical uncertainties which were included in our exploratory analysis, including the choice of asymptotic giant branch yields, the metallicity-resolution of our array of nova ejecta-delay time distributions, common-envelope physics, and nova accretion efficiencies. Further, physical uncertainties appear to be of secondary importance when compared to the large amount of scatter present in observations of \iso{7}Be in nova ejecta. 

\acknowledgments
The authors wish to thank the anonymous referee for their helpful feedback.
A.~R.~C. is supported in part by the Australian Research Council through a Discovery Early Career Researcher Award (DE190100656).
B.~C. acknowledges support from the National Science Foundation (NSF, USA) under grant No. PHY-1430152 (JINA Center for the Evolution of the Elements).
R.~G.~I. thanks the STFC for funding, in particular Rutherford fellowship ST/L003910/1 and consolidated grant ST/R000603/1.
Parts of this research were supported by the Australian Research Council Centre of Excellence for All Sky Astrophysics in 3 Dimensions (ASTRO 3D), through project number CE170100013.

\appendix
\section{Key features of the nova population: ejecta mass}

Most of the key features of our nova populations are discussed in \cite{kemp2021} and \cite{kemp2022}, which discuss which binary systems produce the most novae. However, the key features and distributions of nova properties when considering which systems produce the most nova ejecta requires a brief description.

The relative importance of different WD masses in terms of the number of novae they produce is a balance between the initial mass function, which disfavors high-mass WD systems, and the increased rate at which these systems can produce novae because of their higher surface gravity and lower critical ignition masses (the mass of accreted material required for a nova eruption to occur).

However, reducing the critical ignition mass also reduces the mass of ejecta, which is further complicated by the question of the accretion efficiency (the fraction of accreted material that is retained by the WD beyond the outburst). Increasing the WD mass can lead to higher accretion efficiencies \citep{wang2018}, further reducing the ejecta mass per eruption, but this effect is of secondary importance to the reduced critical ignition mass.

As shown in Figure \ref{fig:dtds}, comparing the ejecta-delay time distributions to their delay time counterparts reveals that late time ejecta-mass contributions are significantly higher than when considering the raw nova rate. This is because the more massive O/Ne WDs contribute far less ejecta per event than their low-mass C/O WD counterparts, reducing the relative importance of their contributions to the distribution.

In general, nova systems which require high (\Mig$\ \gtrsim$ \timestento{5}{-5} M\solar) critical ignition masses for outburst -- typically characterised by low mass WDs (\Mwd$\ < $ 0.8 M\solar) without extremely high accretion rates (\Mdot$\ \lesssim$ \tento{-7} M\solarperyr) -- dominate in terms of the mass of nova processed material. High mass (\Mwd$\ > $ 1 M\solar) WD systems are significantly less important when considering the ejecta mass. However, this does not preclude high mass WDs contributing at a level comparable or even higher than low mass WDs when it comes to the nucleosynthesis of specific isotopes. The higher temperatures obtainable in nova eruptions on massive WDs offer unique nucleosynthetic pathways leading to far more efficient synthesis of elements with mass-numbers higher than F, an effect able to outweigh their lower ejecta masses and system birth rates.

\bibliographystyle{apalike}
\bibliography{bibfile.bib} 

\end{document}